\documentclass[usenatbib]{mnras}

\usepackage{mathptmx}
\usepackage[T1]{fontenc}
\usepackage[british]{babel}

\usepackage[utf8]{inputenc}
\usepackage[T1]{fontenc}
\usepackage{lmodern}

\usepackage{ae,aecompl}

\usepackage{graphicx}	
\usepackage{amsmath}	
\usepackage{amssymb}	

\usepackage{tablefootnote}
\usepackage{savefnmark}
\usepackage{epstopdf}
\usepackage{natbib}
\usepackage{hyperref}

\title[TTVs in Kepler-410Ab]{Transit Timing Variations in the system Kepler-410Ab}

\author[P. Gajdo\v{s} et al.]{Pavol Gajdo\v{s}$^{1}$\thanks{E-mail: pavol.gajdos@student.upjs.sk}, 
\v{S}tefan Parimucha$^{1}$, \v{L}ubom\'ir Hamb\'{a}lek$^2$, Martin Va\v{n}ko$^2$
\\
$^{1}$Institute of Physics, Faculty of Science, Pavol Jozef \v{S}af\'arik University, Ko\v{s}ice, Slovakia\\
$^{2}$Astronomical Institute, Slovak Academy of Sciences, 059 60 Tatransk\'a Lomnica, Slovakia\\}

\date{Accepted 2017 April 17. Received 2017 April 15; in original form 2017 February 24}

\pubyear{2017}

\begin{document}
\setcounter{page}{2907}
\label{firstpage}
\volume{469}
\pagerange{\pageref{firstpage}--\pageref{lastpage}}
\maketitle

\begin{abstract}
We present a new analysis of the transit timing variations displayed by the extrasolar planet Kepler-410Ab. We obtained and
improved orbital and physical parameters for the planet and analysed 70 transit times obtained by the Kepler satellite. 
In our analysis of the O-C diagram (Observed-Calculated), we assumed that the observed changes in the transit times are probably
caused by the gravitational influence of another body in the system. 
To determine the mass of the perturbing body, we have considered the light-time effect
and an analytical approximation of the perturbation model. 
The solutions resulting from both methods give comparable results, with an orbital period \mbox{$P_3\sim$970~days} and 
a slightly eccentric orbit of the third body. We also showed that this orbit is nearly
coplanar with the orbit of the Neptune-like planet Kepler-410Ab (orbital period 17.8 days).
We propose two possible models for the perturbing body orbiting a common barycentre with Kepler-410A:
(i) a single star with mass at least 0.906~M$_{\sun}$,
(ii) a binary star with the total mass of its components of at least 2.15~M$_{\sun}$.
In both cases the star Kepler-410B is on a long orbit (period more than 2200 years).
Small amplitude variations ($\sim$ 5--8 minutes) detected in O-C residuals
can be explained by the stellar activity of the host star (spots and pulsations), which
affects the shape of the light curve during the transit. The presence of single or binary companion of mentioned 
masses heavily affects the total observed flux from the system. After removing of the flux contamination from 
Kepler-410A light curve we found that radius of the transiting planet Kepler-410Ab should be in the range from  about 
3.7 to  4.2 R$_{\earth}$. 
\end{abstract}


\section{Introduction}

The Kepler mission, launched in 2009, has produced photometric data with unprecedented precision \citep{borucki2010}. It was 
designed to detect Earth-size planets orbiting in the habitable zone of parent stars using the transit method. 
The Kepler spacecraft has revolutionized the study of exoplanets, variable stars and
stellar astrophysics; providing photometric data with high-precision, high-cadence continuous light curves
(LCs). After losing two reaction wheels, the Kepler spacecraft ended its primary mission and subsequently started its K2 mission started 
\citep{howell2014}. The photometric precision of K2 is slightly lower, but still much better than that from ground-based 
observatories. 

The high precision of Kepler data and the long-term uninterrupted observations enable us to determine not only transit 
parameters, but also to measure small changes in the individual times of transits with respect to a linear ephemeris that 
assumes a Keplerian orbit. These variations (often called Transit-Timing Variations, TTVs) can reflect dynamical 
interaction with other objects in the system (other star(s), planets(s)) which cannot be discovered directly by the transit method. 
Gravitational interactions between the non-transiting object and the transiting planet can cause a periodic shift of the star-planet barycentre which can be detected
through changes in timings of transits (so called light-time effect). Alternatively, TTVs can also be 
caused by stellar activity when a surface spot deforms the shape of the transit which leads to systematic shifts in transit timing.  

Up to now (24 February 2017), 2330 confirmed planets and 4706 planet candidates have been discovered by Kepler.
Another 173 planets and 458 candidates have been detected during the K2 mission. 
One of the confirmed transiting exoplanets is \mbox{Kepler-410Ab}, a Neptune-sized planet on 17.8336 days orbit, 
discovered in 2013 and confirmed by \citet{Eylen}. The host 
star \mbox{Kepler-410A} (KIC 8866102, KOI-42, HD 175289) is a young \mbox{$2.76 \pm 0.54$~Gyr} old star with radius \mbox{$1.352 \pm 0.010$~R$_{\sun}$}
and mass \mbox{$1.214 \pm 0.033$~M$_{\sun}$} and a spectral type F6IV \citep{Molenda}. The first GAIA Data Release
\citep{brown2016} gives a distance of \mbox{$153.6^{+6.7}_{-6.1}$~pc}.

Using adaptive optics, \citet{Kep410B} distinguished a small stellar companion Kepler-410B separated by an angular distance
of $1''.63$. \cite{Eylen} found that this star is probably a red dwarf and ruled it out as a host star for the exoplanet.

TTVs of Kepler-410Ab were for the first time reported by \citet{Mazeh}. They studied the TTVs using a sinusoidal model 
and found an O-C semi-amplitude of \mbox{$13.91 \pm 0.91$} minutes and a period of about 960 days. In a subsequent analysis, \citet{Eylen}
used a 'zigzag' model. They obtained a slightly larger semi-amplitude of 16.5 minutes, and a period of 957 days. They also noted that the 
shape of the O-C in not sinusoidal, which could be caused by an eccentric orbit of Kepler-410Ab.

In this paper we give a new analysis of the TTVs in the orbit of Kepler-410Ab. In Section \ref{time}, we describe the determination of 
the parameters of the exoplanet and the individual transit times of Kepler-410Ab. In the next Section \ref{oc}, 
we provide the physical models used for fitting TTVs. Our results are discussed in Section \ref{discusion}.

\begin{figure}
\includegraphics[width=0.49\textwidth]{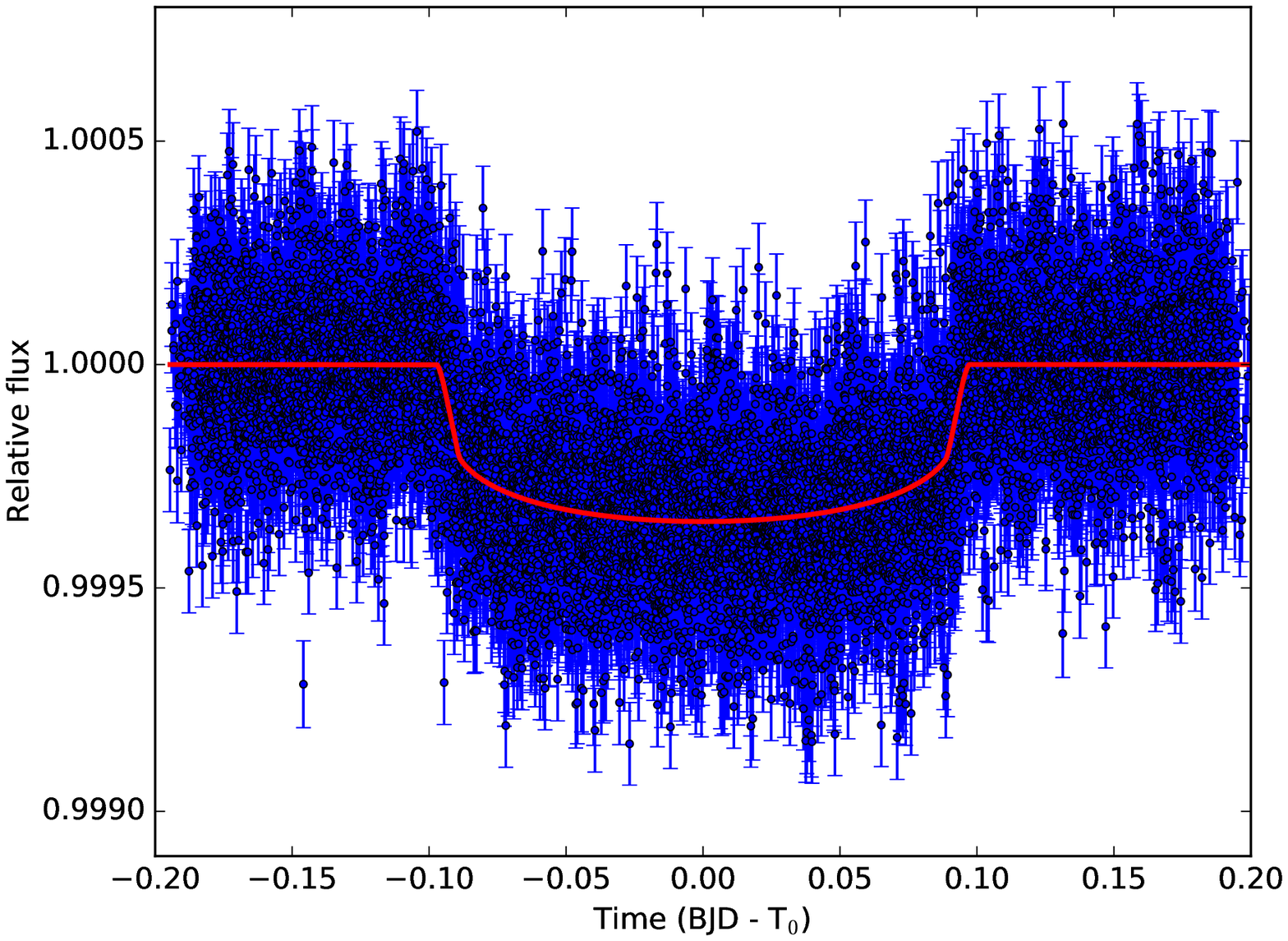}
\includegraphics[width=0.49\textwidth]{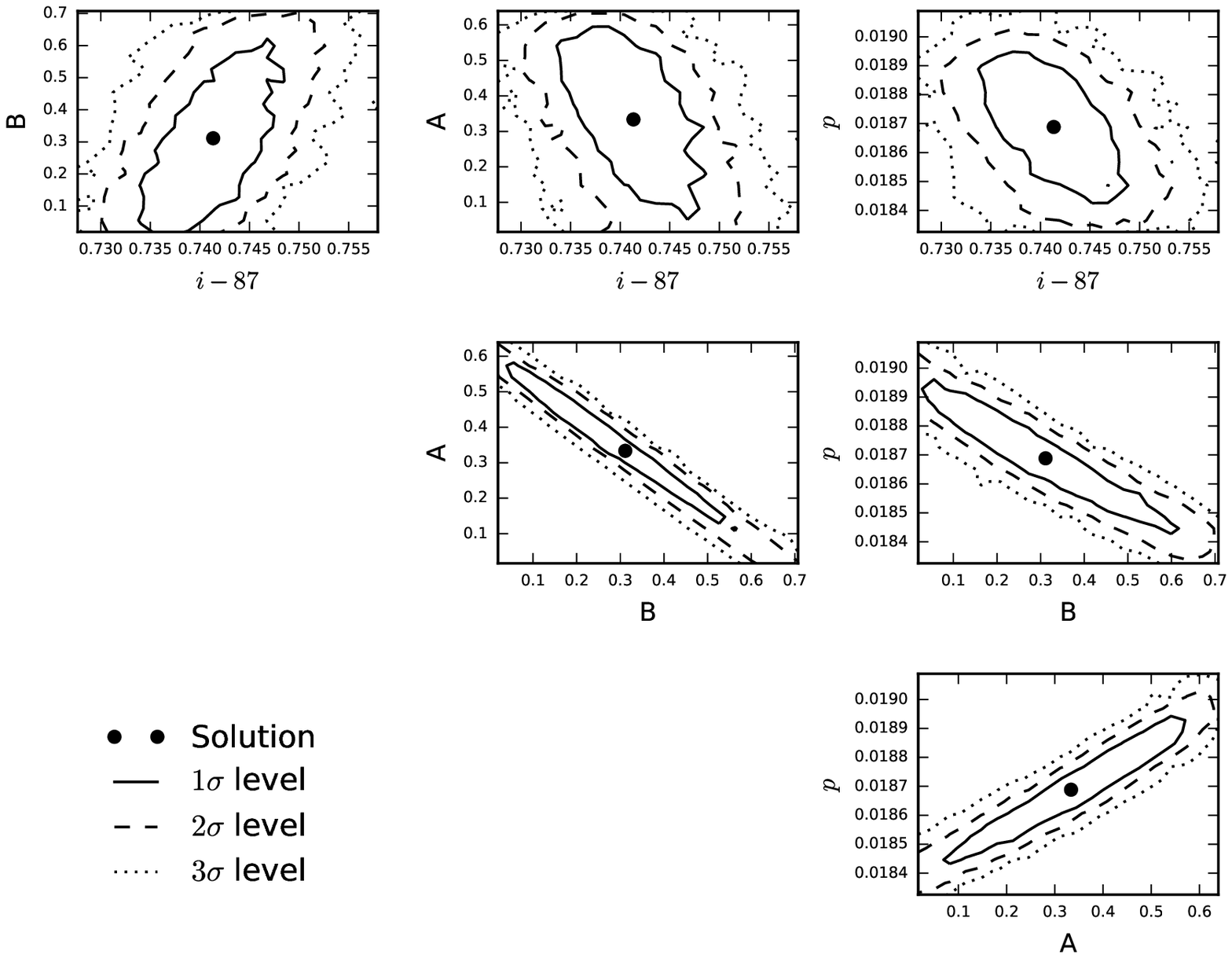}
\caption{(top) -- Stacked transits of Kepler-410Ab folded by the procedure described in Section \ref{time}. 
Original Kepler data are shown as blue points with error bars, the red line marks the best fit of the transit model
with parameters in Tab. \ref{tab:parameters}. 
(bottom) -- Confidence interval graphs for fitted parameters (orbital inclination $i$, planet-star size ratio $p$ and 
limb darkening coefficients $A$ and $B$) determined by MCMC simulation.}
\label{fig:trans}
\end{figure}

\begin{table*}
\caption{Transit parameters of exoplanet Kepler-410A determined for different flux contaminations of 
Kepler-410A light curve (see text), $a$ - semi-major axis of the planetary orbit \citep[adopted from][]{Eylen}, $r_{\rm 
p}$ - planet radius, $i$ - orbital inclination, $A$, $B$ - linear and quadratic coefficients of the 
limb-darkening, $\chi^2$ - sum of squares of the best fit, $\chi^2/n$ - reduced sum of squares.}
\label{tab:parameters}
\begin{center}
\begin{tabular}{lccc}
\hline
Parameter          		& Solution 1 		& Solution  2		  & Solution 3 		 	   \\
\hline
Flux contamination		& 8\%      		& 42\%      	  	  & 53\% 	    \\
				& Kepler-410B	& Kepler-410B+star 	  & Kepler-410B+binary    \\
\hline
$a$ [au] 		        & \multicolumn {3}{c}{ 0.1226$\pm$0.0047 $^a$} \\
$a$ [$R_\star$] 	   	& \multicolumn {3}{c}{19.53$\pm$0.76 $^a$ }\\
$r_{\rm p}$ [$R_{\earth}$] 	& 2.647$\pm$0.020       & 3.744$\pm$0.029        & 4.239$\pm$0.033            \\
$r_{\rm p}$ [$R_\star$] 	& 0.01798$\pm$0.00004   & 0.02543$\pm$0.00006    & 0.02879$\pm$0.00006        \\
$i$ [\degr] 		     	& 87.744$\pm$0.003       & 87.745$\pm$0.003       & 87.745$\pm$0.003          \\
$A$ 			     	& 0.3118$\pm$0.0404       & 0.3157$\pm$0.0421       & 0.3178$\pm$0.0450       \\
$B$ 	 		     	& 0.3310$\pm$0.0457       & 0.3278$\pm$0.0488      & 0.3269$\pm$0.0492        \\
\hline                                                                                                                              
$\chi^2$ 		        & 33181.59               & 33214.11               & 33226.79                  \\
$\chi^2/n$ 		     	& 3.01                   & 3.02                   & 3.02                      \\
\hline
$^a$ adopted fixed value
\end{tabular}
\end{center}
\end{table*}

\section{Transit times determination}
\label{time}

For the determination of the individual times of transit we used short-cadence (sampled every 58.8 seconds) de-trended 
data (PDCSAP\_FLUX) from quarters Q1 to Q17, provided by the NASA Exoplanet 
Archive\footnote{\href{http://exoplanetarchive.ipac.caltech.edu}{http://exoplanetarchive.ipac.caltech.edu}}.

As a first step, we extracted parts of the LC around detected transits using the ephemeris given in \cite{Eylen},
where we took an interval $\pm$0.2 days around the computed transit time $T_{\rm T}$ (the interval size is approximately
double the transit duration).  To remove additional residual trends caused by the stellar activity and instrumental long-term photometric
variation, we fitted the out-of-transit part of LC by a second-order polynomial function. 
Then we subtracted 8\% flux contamination from the wide companion Kepler-410B, according to calculations of 
\citet{Eylen}
All individual parts of the LC with transits were stacked together to obtain the template of the transit.
This can be done, because one expects that the physical parameters of the host star
and the exoplanet did not change during the observational period of about 3.5 years and we want to cancel-out the 
effect of starspots.

The stacked LC was fitted by our software implementation of \citet{MA} model, where we used the
Markov Chain Monte Carlo (MCMC) simulation method for the determination of transit parameters. This method takes into account 
individual errors of Kepler observations and gives a realistic and statistically significant estimate of parameter 
errors. As a starting point for the MCMC fitting, we used the physical parameters of the planet given in 
\citet{Eylen}. Because of the strong correlation between the orbital inclination $i$ and the semi-major axis $a$ of the 
planetary orbit, we have adopted a fixed value \mbox{$a=0.1226\pm0.0047$~au}.
We have used a quadratic model of limb darkening with starting values of coefficients (linear term $A$ and quadratic 
term $B$) from \citet{Sing}.
We ran the MCMC simulation with 10$^6$ steps. The convergence of MCMC fit was checked using the Geweke diagnostic 
\citep{geweke}.

We have repeated the MCMC simulation with the previous solution as the starting point on each of 70 individual transit intervals, and let only
the time of transit $T_{\rm T}$ to update. The new values of $T_{\rm T}$ were used to improve the linear ephemeris and to 
construct a new O-C diagram.

The combined light curve stacked using a linear ephemeris is affected by relatively large amplitude of O-C time shifts. To correct 
this effect, we used iterative procedure that takes the best-fit O-C values (see Section \ref{oc}) into account. 
Afterwards, a new stacked light curve was constructed and a new MCMC transit solution was calculated, subsequently a new 
ephemeris and O-C values were determined. This process was repeated three times until a convergent solution was 
reached. 

The resulting values of fitted transit parameter are listed in Tab. \ref{tab:parameters} (Solution 1). Corrected 
stacked transit light curve with the best fit is displayed in Fig. \ref{fig:trans} (top) and confidence intervals of 
fitted parameters are displayed in Fig. \ref{fig:trans} (bottom).

A new improved linear ephemeris was determined using obtained values of $T_T$ (Tab. \ref{tab:transits}):
\begin{equation}
\label{ephemeris}
T_{\rm T} = 2455014.23765 (22)+17.8336313 (43) \times E,
\end{equation}
where $E$ is the epoch of observation.
Using the improved ephemeris (\ref{ephemeris}), the corrected O-C diagram was constructed (Fig. \ref{fig:oc}).

\section{Analysis of O-C diagram}
\label{oc}

\begin{table}
\caption{Barycentric transit times $T_{\rm T}$ of Kepler-410Ab, with their uncertainties $\sigma_{T_{\rm T}}$,
$\chi^2$ and $\chi^2/n$ statistics ($n$ - the number of data points in the fit). 
The full table is available as a supplementary material to this manuscript.}
\label{tab:transits}
\begin{center}
\begin{tabular}{ccccc}
\hline
$T_{\rm T}$ [BJD]   & $\sigma_{T_{\rm T}}$ & $\chi^2$ & $\chi^2/n$ & $n$\\ 
\hline       
2454978.56473 & 0.00063        & 647.713  & 1.186        & 556 \\
2454996.39755 & 0.00066        & 637.994  & 1.168        & 556 \\
2455103.39002 & 0.00059        & 2509.821 & 4.709        & 543 \\
2455121.22740 & 0.00069        & 2441.934 & 4.581        & 543 \\
\dots  	      & \dots          & \dots    & \dots        & \dots \\
\hline
\end{tabular}
\end{center}
\end{table}

Our transit times $T_{\rm T}$ show a periodic variation with an amplitude of approximately 30 minutes and a period between 950 and 1000 days, which is in agreement with previous \mbox{analyses} by \citet{Mazeh} and \citet{Eylen}. 

Let us denote a linear ephemeris
\begin{equation}\noindent
\label{eq:linear}
T_{\rm C} = t_{\rm 0} + P \times E,
\end{equation}
where $t_{\rm 0}$ is the initial time of transit, $P$ is the orbital period of the planet, and $E$ is the epoch 
of observation. The observed transit time $T_{\rm T}$ can be calculated by adding a perturbation $\delta T$
to the linear ephemeris (\ref{eq:linear}):
\begin{equation}
T_{\rm T} = T_{\rm C} + \delta T.
\end{equation}

Assuming that the observed TTVs are due to the gravitational influence of another body (planet or star) in the system, we 
can calculate $\delta T$ to find the physical parameters of the perturbing body using two different approaches:

\begin{enumerate}
\item LiTE (light-time effect) solution \citep{irwin1952}. This method is often
used to find an unseen companion in binary stars.
A perturbation $\delta T$ caused by the third body is described by the equation:
\begin{equation}
\label{eq:lite}
\delta T = \frac{a_{12}\sin i_3}{c}\left[\frac{1-e_{3}^2}{1+e_3\cos\nu_3}\sin(\nu_3+\omega_3)+e_3\sin\omega_3\right], 	
\end{equation}
where $c$ is the speed of light, $a_{12}$ is semi-major axis of the orbit of Kepler-410A, and $i_3$, $e_3$, $\omega_3$, and $\nu_3$ are, respectively, 
the inclination, the eccentricity, the argument of the periastron,
and the true anomaly of the orbit of the third-body around the barycentre of the system.
Since the value of $i_3$ is not known, we can determine only the mass function
for the third body:
\begin{equation}
\label{eq:massfun}
f(M_3) = \frac{(M_3 \sin i_3)^3}{M^2} = \frac{(a_{12}\sin i_3)^3}{P_3^2},
\end{equation}
where $M = M_{\star} + m + M_3$ is the total mass of the system
(sum of the masses of the host star, the planet, and the third body, respectively).
The minimal mass $M_3$ can be calculated by assuming a coplanar orbit, i.e. $i_3 = i = 87.\degr74$.
The period of the third body $P_3$ and the time of pericenter passage $t_{03}$
are hidden in the $\nu_3$ calculation and need to be solved for using the Kepler's equation. \\

\item an analytical approximation of the perturbation model given in \citet{Agol}.
This method was originally developed for the analysis of multiple planetary systems,
but is not constrained to any limit of mass of the perturbing body.
An exterior body on a large eccentric coplanar orbit causes a time variation (eq. 25 in \citet{Agol}):
\begin{equation}
\label{eq:agol}
\begin{aligned}		
\delta T = & \frac{M_3}{2\pi(M_\star+m)} \frac{P^2}{P_3} (1-e_3^2)^{-3/2} \\ &
	      \left[\nu_3-n_3(t-t_{03})+e_3\sin \nu_3 \right],
\end{aligned}
\end{equation}
where $n_3 = 2\pi/P_3$ is the mean motion. 
We can reduce the number of parameters in the model by substituting
\begin{equation}
\label{eq:mass}
\frac{M_3}{M_\star + m} = \frac{\mu_3}{1-\mu_3} \ ,
\end{equation}
where $\mu_3 = M_3/M$ is the reduced mass of the third body. Because the
O-C amplitude is large, we expect the mass of the perturbing body to be of the order
of one solar mass. Therefore, we can neglect the mass of the planet $m\ll M_3$.
\end{enumerate}

\begin{figure}
\includegraphics[width=0.49\textwidth]{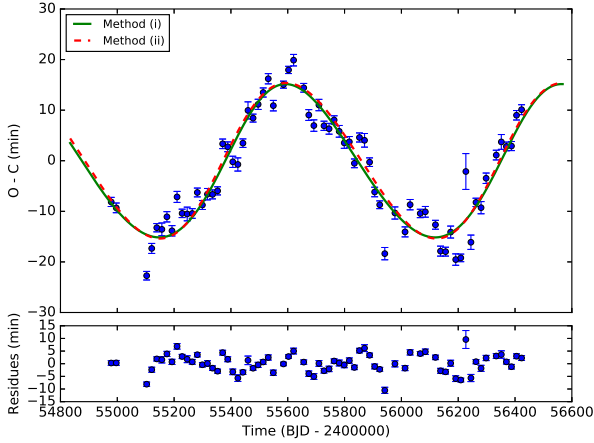}
\caption{The O-C diagram of Kepler-410Ab constructed according to the improved linear ephemeris \eqref{ephemeris}.
The solid green line represents the best fit computed from the LiTE model, while
the dashed red line corresponds to the solution according to the perturbation model by \citet{Agol}.
The residuals are plotted against the statistically better solution by the LiTE model.}
\label{fig:oc}
\end{figure}

\begin{figure}
\includegraphics[width=0.49\textwidth]{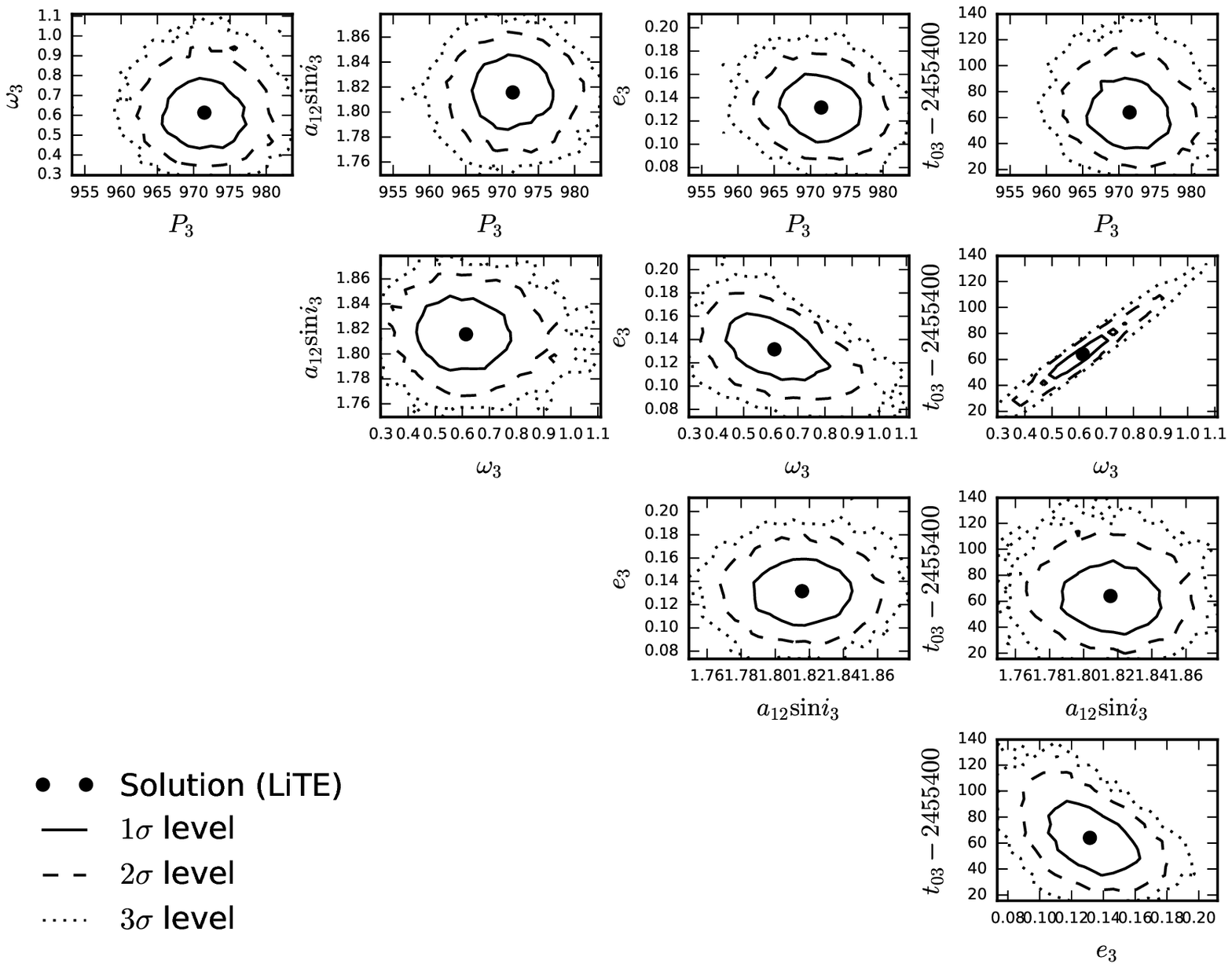}
\includegraphics[width=0.49\textwidth]{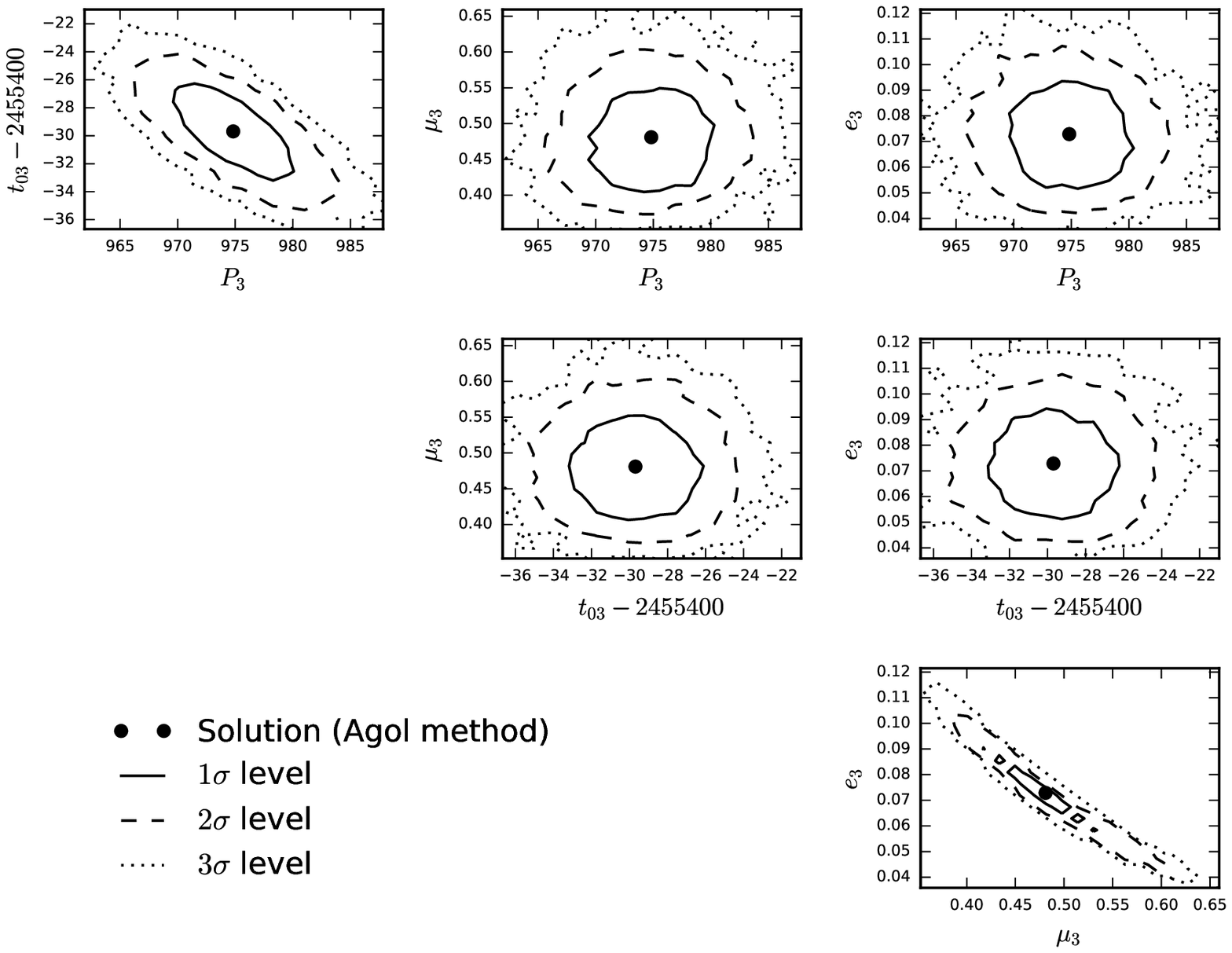}
\caption{Confidence intervals determined by the MCMC simulation for parameters fitted in both methods.}  
\label{fig:oc-conf}
\end{figure}

To obtain the optimal parameters with statistically significant errors in both approaches, we have used our own code 
based on genetic algorithms for the determination of the initial values of all parameters, and MCMC simulation for the
final solution and error estimation. The uncertainty of each individual $T_{\rm T}$ were considered. A full description of 
our code is in preparation \citep{OCSolver}. The results from both methods are listed in Tab. \ref{tab:3body} and 
depicted in Fig. \ref{fig:oc} with confidence intervals shown in Fig. \ref{fig:oc-conf}.
To estimate the quality of the statistical model, we have also calculated the Bayesian 
information criterion (BIC) defined as:
\begin{equation*}
\mathrm{BIC} = \chi^2 + g \ln n
\end{equation*}
where $g$ is the number of degrees of freedom, and $n$ is the number of data points (TTVs). The lower 
BIC score means that a model is statistically more significant.

\section{Discussion and conclusion}
\label{discusion}
Previous studies by \citet{Mazeh} and \citet{Eylen} dealt with only a quantitative analysis of TTVs in Kepler-410A system.
The nature of the perturbing body was not discussed. 

Our interpretation is based on a natural assumption that TTVs are caused by a gravitational influence of another body in the system. This is similar to period variations caused by a third body observed in eclipsing binaries.

It is obvious that a period of $\sim$970 days in the $O-C$ diagram cannot be caused by the companion star Kepler-410B. If we assume that this star is a low-mass
(0.5-0.8~M$_{\sun}$) cool star, based on the spectral type corresponding to the temperature $\sim4850$~K derived by \citet{Eylen};
and we take into account the observed separation ($1''.63$) and the distance to 
system (153.6 pc), we can calculate the orbital period of Kepler-410B to be in the range of 2200--2500 years.

The fact that the shape of the O-C diagram is not strictly sinusoidal (as first noted by \citet{Eylen})
agrees with our modelled eccentricity $e_3$ of the orbit of the third body.
The orbital inclination $i_3$ cannot be determined directly from the O-C diagram. The analytical approximation
of the perturbation model \citep{Agol} assumes that the orbit is coplanar with that of the exoplanet and the planet is seen edge-on.
These assumptions are valid, because the derived inclination of the planet is $i=87.\degr74$.

To calculate the mass of the third body, we have neglected the mass of the planet $m\ll M_3$.
According to the measured radius in the interval from $R_p=2.65~R_{\earth}$ to $R_p=4.24~R_{\earth}$ (see 
Tab.~\ref{tab:parameters}) the exoplanet Kepler-410Ab is a Neptune-like body and we can assume its mass of the order of 
several Neptune masses 
\mbox{$m\sim10^{-4}$~M$_{\sun}$}. The amplitude of the O-C curve is $\sim$30 minutes which is too high to be caused by a body 
with sub-stellar mass. The resulting mass of the third body \mbox{$M_3=0.906\pm0.155$~M$_{\sun}$} should be regarded as a 
lower limit, because of the assumptions.

\begin{table}
\caption{Parameters of the third body obtained from two methods described in Section \ref{oc}.} 
\label{tab:3body}
\begin{center}
\begin{tabular}{lcc}
\hline
Parameter     		    & LiTE solution 	         & Agol method \\
\hline
$P_3$ [days]		      & 971.1$\pm$3.7		         & 973.6$\pm$3.6\\
$e_3$			            & 0.15$\pm$0.02		         & 0.09$\pm$0.01\\
$t_{03}$ [BJD]		    & 2455440.2428$\pm$15.5311 & 2455372.1803$\pm$2.2120\\
$a_{12}\sin i_3$ [au]	  & 1.839$\pm$0.020 		     & --\\
$\omega_3$ [\degr]	  & 25.8$\pm$5.8		         & --\\
$\mu_3$			          & --			                 & 0.428$\pm$0.041 \\
$f(M_3)$ [M$_{\sun}$]	& 0.879$\pm$0.030	         & --\\
$M_3$ [M$_{\sun}$]	  & 2.151$\pm$0.078     	   & 0.906$\pm$0.155\\
\hline
$\chi^2$ 		          & 804.9	                   & 831.5 \\ 
$\chi^2/n$ 		      & 12.4	                   & 12.6 \\ 
BIC			              & 826.2	                   & 848.5 \\ 
\hline
\end{tabular}
\end{center}
\end{table}

Using the adopted LiTE solution, we arrived at the mass function \eqref{eq:massfun} 
\mbox{$f(M_3)=0.879\pm0.030$~M$_{\sun}$}. For a coplanar orbit (substituting $i$ for $i_3$) we get the minimal 
mass of the body \mbox{$M_3=2.151\pm0.078$~M$_{\sun}$}. If we consider a main-sequence star, this would correspond to the spectral type A4
(with \mbox{$T_{\rm eff}\sim8500$~K}).
But according to spectra obtained by \citet{Molenda}, the host star Kepler-410A is classified as F6IV.
Moreover, during the $\sim$1460-days long observation by Kepler, no signs of eclipses with $\sim$970 days period were found.
The mutual distance of bodies with masses \mbox{$M_{\star}=1.2$~M$_{\sun}$} and \mbox{$M_3=2.1$~M$_{\sun}$}
orbiting with the period of $\sim$970 days is about 615~R$_{\sun}$. For stars with radii
\mbox{$R_{\star}=1.35$~R$_{\sun}$} \citep{Eylen} and \mbox{$R_3\sim1.8$~R$_{\sun}$} (according to assumed spectral type A4) eclipses will start to occur
only if the inclination \mbox{$i_3>89.\degr87$}, which explains the non-detection of additional transits.
A more inclined orbit leads to more massive star with greater luminosity. This would favour the spectroscopic detection.
However, if the perturbing body is comprised of a non-eclipsing binary star, the combined luminosity of
both stellar components could still be under the detection limit of the spectrograph used by \citet{Molenda}.

\begin{figure}
\includegraphics[width=0.49\textwidth]{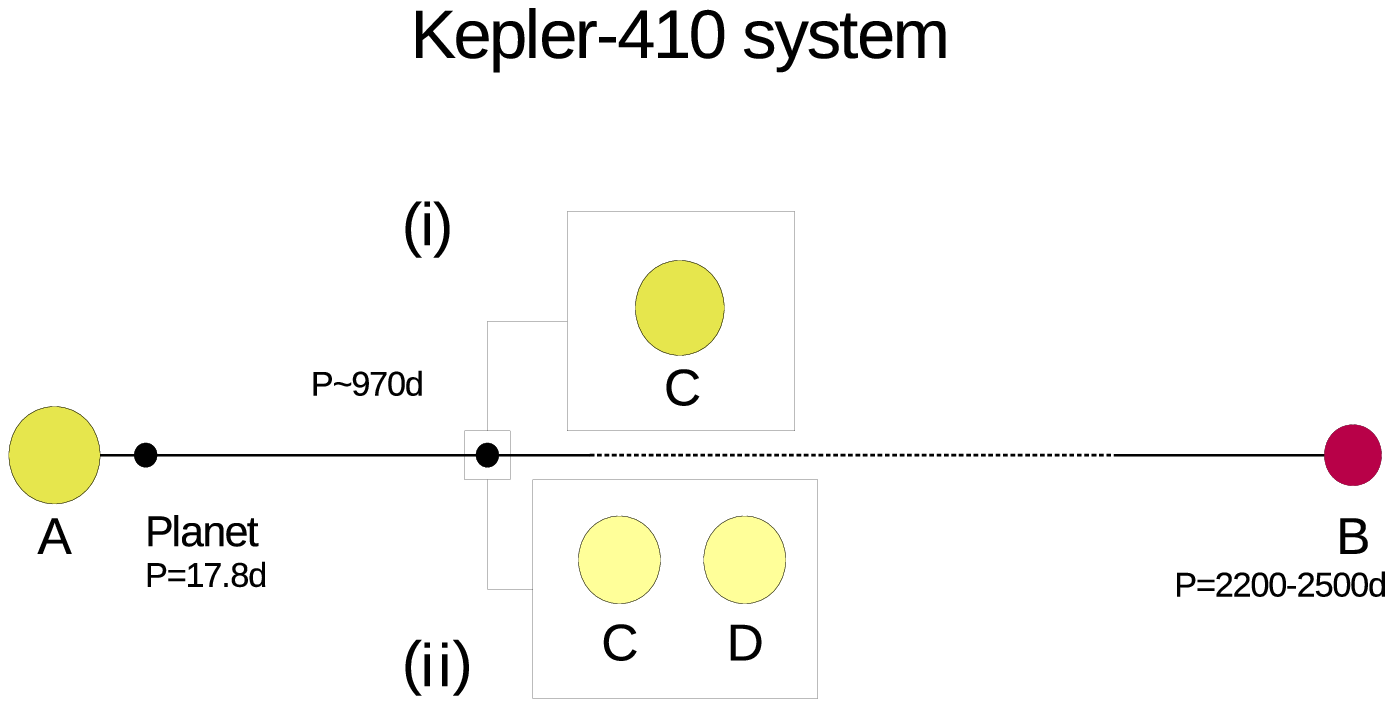}
\vspace{-1cm}
\caption{Schematic view of Kepler-410 system (not in scale). Planet b is orbiting the host star Kepler-410A with 
period 17.8 days, the orbital period of the companion star Kepler-410B is about 2200--2500 years,
and the perturbing component (with orbital period relative to the system barycentre $\sim$970 days) could be:
(i) a single star with minimal mass 1.12~M$_{\sun}$, or
(ii) a close binary star with the total mass of components at least 2.1~M$_{\sun}$.} 
\label{fig:model}
\end{figure}

The light curve of Kepler-410A is affected by rotational flux modulation with the amplitude 7--8 mmag and 
the period about 25 days. The amplitude of exoplanet transits is only $\sim$4 mmag.
To remove the signal of the modulation, we selected parts of LC with transits shorter than $\sim$0.2 days
and fitted the out-of-transit part (see Section~\ref{time}). 
The amplitude of O-C constructed with the updated ephemeris is about 30 minutes, which agrees with previous estimate
by \citet{Eylen} and \citet{Mazeh}. Some low-amplitude variations with periods in range of $\sim$ 200 days are 
still visible in residuals (Fig. \ref{fig:oc}). These can be interpreted by the presence of surface spots \citep{Mazeh2015}.

We propose the following scenarios to explain the observed TTVs in transits of Kepler-410A (see Fig. \ref{fig:model}).
The perturbing body is:
\begin{enumerate}
\item a star with minimum mass of 0.91~M$_{\sun}$ in orbit around the common barycentre with Kepler-410A and the period $\sim$970 days.
\item	a non-eclipsing binary star with minimum total mass of components 2.15~M$_{\sun}$. The binary forms a hierarchical
system with Kepler-410A.
\end{enumerate}
In both cases, the component Kepler-410B is a cool red dwarf star on a distant orbit
with period more than 2200 years and can not be the originator of observed TTVs.

We made simple N-body simulations of this problem for edge-on orbits. We used Everhart's improvements of Gauss-Radau integrator \citep{Everhart1985}. 
Obtained O-Cs are very similar for both porposed scenarios. 
Any significant difference due to a possible binary character of perturbing compunent was not observed. 
The shape of model O-Cs (Fig.~\ref{fig:oc}) was reproduced in our simulations. 
Based on our simulations, we could not explain the variations in the residuals to the O-C diagram.
The oscillatory features of the residuals are not caused by possible binary character of perturbing compunent.

In both proposed scenarios we found different amounts of flux contamination by the perturbing body during the observed 
transits. If not corrected, the transit depth would be underestimated and the fit parameters would not accurate.
We calculated possible minimum flux contamination using the mass-luminosity relation, from a single main sequence star 
with mass 0.91~M$_{\sun}$ (Solution 2), and from a binary consisting of two main sequence stars with mass of each 
companion about 1M$_{\sun}$ (Solution 3). The calculated contaminations were subtracted from Kepler-410A flux already 
corrected for a 8\% Kepler-410B contribution \citep{Eylen} and a new transit parameters were calculated using MCMC 
method (see Section \ref{time}). The results are listed in Tab. \ref{tab:parameters}. Considering our proposed 
scenarios, the radius of the transiting planet Kepler-410Ab should be in the range from about $R_p$ = 3.74 R$_{\earth}$ to  
$R_p$=4.24 R$_{\earth}$. 
These values are significantly larger than that determined by \citet{Eylen}. Other transit parameters as well as transit times were not affected by the correction to flux contamination.

Surprisingly, no systematic spectroscopic observations of this system are available
to this date. In 2016, we started a low-dispersion spectroscopic campaign using a 60-cm telescope 
\citep{Pribulla2015}. Preliminary results suggest radial-velocity variations of Kepler-410A.
For a definitive answer, we plan an observation program for high-dispersion spectroscopy of Kepler-410.
Using the broadening-function method \citep{Rucinski2002}, we want to investigate the clues
of radial velocities of any potential additional source.

\section*{Acknowledgement}
We would like to thank the anonymous referee for helpful comments and
corrections. This paper has been supported by the grant of the Slovak Research and Development Agency with number 
APVV-15-0458. This article was created by the realization of the project ITMS No.26220120029, based on the supporting 
operational Research and development program financed from the European Regional Development Fund.
\v{L}.H. and M.V. would like to thank the project VEGA 2/0143/14.

\label{lastpage}

\section*{Supporting information}
Additional Supporting Information may be found in the on-line version of this article:\\

\noindent Table \ref{tab:transits}. {Barycentric transit times $T_{\rm T}$ of Kepler-410Ab, with their uncertainties 
$\sigma_{T_{\rm T}}$, $\chi^2$ and $\chi^2/n$ statistics ($n$ - the number of data points in the fit).\\

\noindent Please note: Oxford University Press is not responsible for the content or functionality of any supporting 
materials supplied by the authors. Any queries (other than missing material) should be directed to the corresponding 
author for the paper.
\end{document}